\documentclass{article}

\usepackage{arxiv}

\usepackage[utf8]{inputenc} 
\usepackage[T1]{fontenc}    
\usepackage{hyperref}       
\usepackage{url}            
\usepackage{booktabs}       
\usepackage{amsfonts}       
\usepackage{nicefrac}       
\usepackage{microtype}      
\usepackage{lipsum}

\usepackage{graphicx}
\usepackage{dcolumn}
\usepackage{bm}
\usepackage{amsmath}
\usepackage{subfigure}
\usepackage{epstopdf}
\usepackage{csvsimple,longtable,booktabs}
\usepackage{array}
\usepackage{tabularx}
\newcolumntype{C}[1]{>{\centering\let\newline\\\arraybackslash\hspace{0pt}}m{#1}}

\usepackage{hyperref}
\usepackage[mathlines]{lineno}
\usepackage{color,soul}

\usepackage{enumerate}
\usepackage{enumitem}
\usepackage{newlfont}

\title{Structure dynamics of evolving scientific networks}

\author{
  Demival Vasques Filho \\
  Leibniz-Institut f\"ur Europ\"aische Geschichte,
  Alte Universit\"atsstra{\ss}e 19, 55116 Mainz, Germany \\
  Te P\={u}naha Matatini, Department of Physics, 
  University of Auckland, 
  Private Bag 92019, Auckland, New Zealand \\
  \texttt{vasquesfilho@ieg-mainz.de} \\
   \And
 Dion R.J. O'Neale \\
	Te P\={u}naha Matatini, Department of Physics, University of Auckland, 
	Private Bag 92019, Auckland, New Zealand \\
  \texttt{d.oneale@auckland.ac.nz} \\
}

\begin{document}
\maketitle

\begin{abstract}
Co-authorship networks have been extensively studied in network science as they pose as a perfect example of how single elements of a system give rise to collective phenomena on an intricate, non-trivial structure of interactions. However, co-authorship networks are, in fact, one-mode projections of original bipartite networks, which we call here scientific networks, where authors are agents connected to artifacts - the papers they have published. Nonetheless, few studies take into account the structure of the original bipartite network to understand and explain the topological properties of the projected network. Here, we create bipartite networks using extensive datasets from the American Physical Society (APS) dating back to 1893 up to 2015 inclusive, and from arXiv (1986-2015). We look at the time evolution of publications and at the dynamic structure of scientific networks considering four major features of bipartite networks, namely degree distributions, density, redundancy and cycles. We show how such features shape the formation of co-authorship networks and their observed structural properties. While the structure of the networks of most disciplines does not show significant changes over time, the appearance of large collaborations, in physics, generate largely skewed degree distributions of top nodes. The latter, in turn, induces massive cliques in the projection, triggering considerable densification of the co-authorship network, while the density of the original bipartite network remains at the same levels.
\end{abstract}

\keywords{Bipartite networks \and Co-auhtorship networks \and Degree distribution \and Degree assortativity \and Network cycles}

\section{\label{sec:introduction}Introduction}

Co-authorship (collaboration) networks are among the many real-world networks that have an underlying bipartite structure. In this case, the bipartite network $B$ has the bottom set of nodes $U$ composed of authors and the top set $V$ composed of the papers that these authors published in scientific journals. The co-authorship network is then the one-mode network $G$ created via a projection onto the bottom nodes (authors). Authors are connected if they share a common neighbour (paper) in $B$ or, in other words, authors are connected in the projected network if they co-author a paper. 

There are a few different ways of creating projections of bipartite networks, for instance, simple graph, multigraph and weighted graph projections \cite{vasques2018degree}. Here, we work with these three methods as we want to keep track of (a) node degree (the number of unique collaborators of each author), (b) link weight (the number of shared publications between any pair of authors), and (c) node strength (the sum of shared publications over the unique collaborators of each author). For the link weight, we use simple-weighted projections which means that the weight of the link connecting two collaborators will be equal to the number of shared publications between them.

There exist a large number of previous studies on co-authorship networks \cite{newman2001structure,acedo2006co,abbasi2011identifying,mali2012dynamic,wagner2005network,glanzel2004analysing,liu2005co}. However, to the best of our knowledge, none of these take in account the structure of the original bipartite network to understand and explain topological properties of the projected network. Here, we create bipartite networks using an extensive database from the American Physical Society (APS) \cite{apsdata} dating from 1893 to 2015 inclusive. Previous works have explored this dataset before, but mainly on citation data \cite{martin2013coauthorship,redner2004citation,chen2010community,radicchi2009diffusion,gualdi2011influence}, again leaving aside the topological properties of both bipartite and projected networks. We also use data collected from the electronic repository of scientific preprints ArXiv (from 1986 to 2015) \cite{arxiv}. We divide the ArXiv dataset into five different networks for each of the available disciplines: physics, mathematics, computer science, biology, and statistics.

\begin{figure*}[]
	\centering
	\subfigure{\label{fig:aps_size} \includegraphics[scale=0.40]{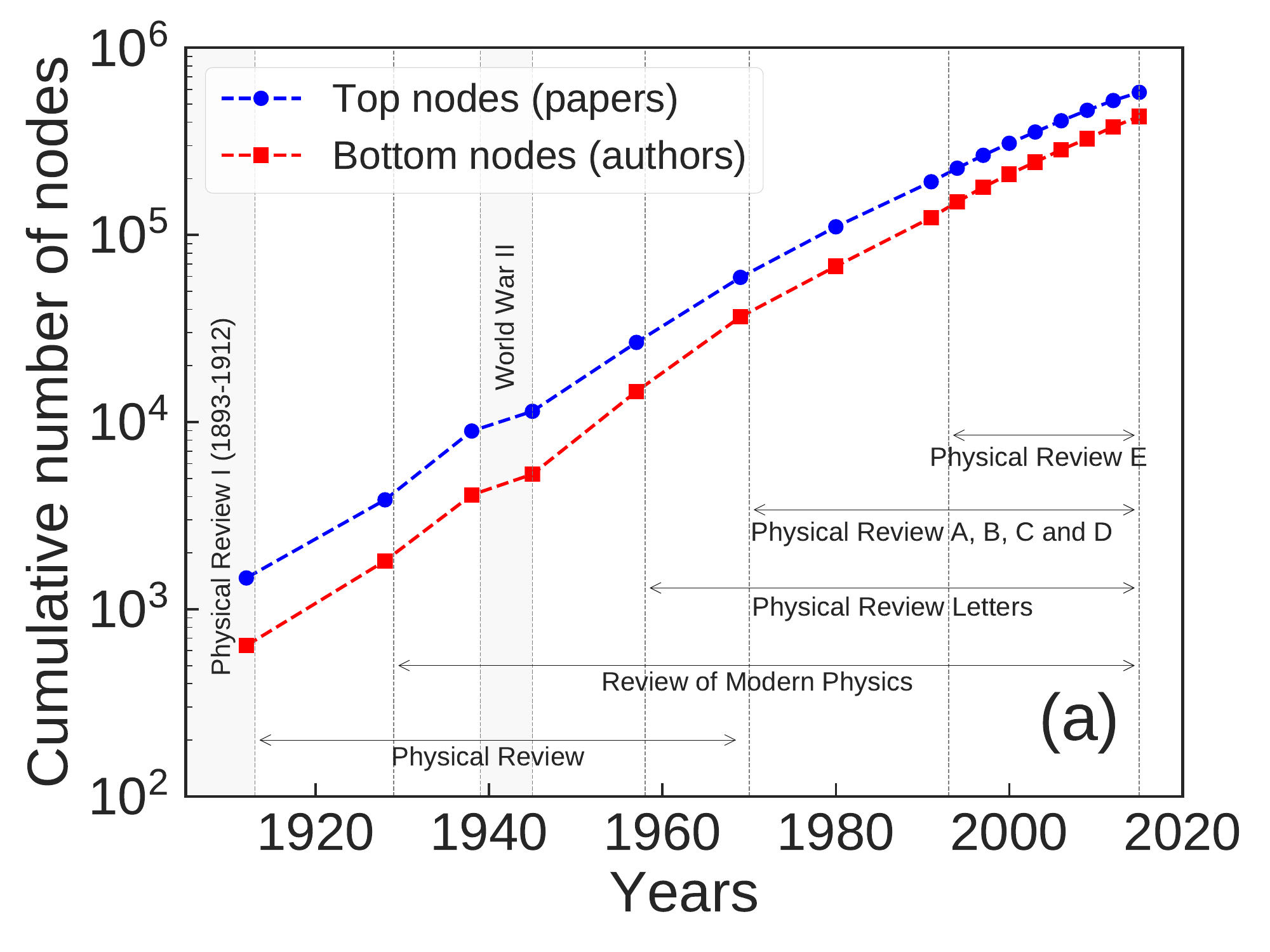}}
	\subfigure{\label{fig:arxiv_size} \includegraphics[scale=0.40]{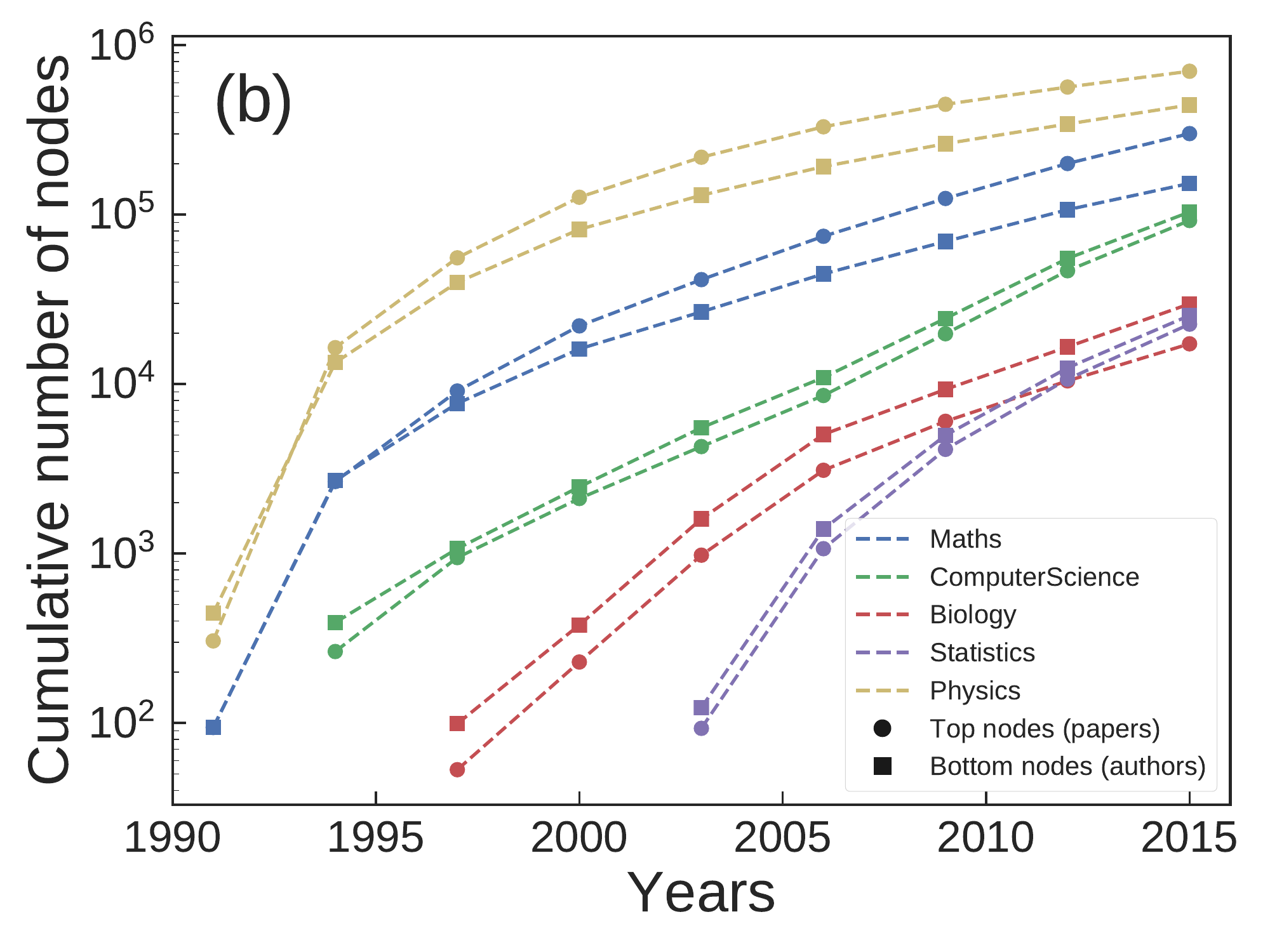}}%
	\caption{Evolution of the size of bipartite networks for (a) the APS journals and (b) ArXiv for the disciplines of physics, maths, computer science, biology and statistics. We can see an exponential growth of all networks.}
	\label{fig:network_size}
\end{figure*}

\begin{figure*}[]
	\includegraphics[scale=0.40]{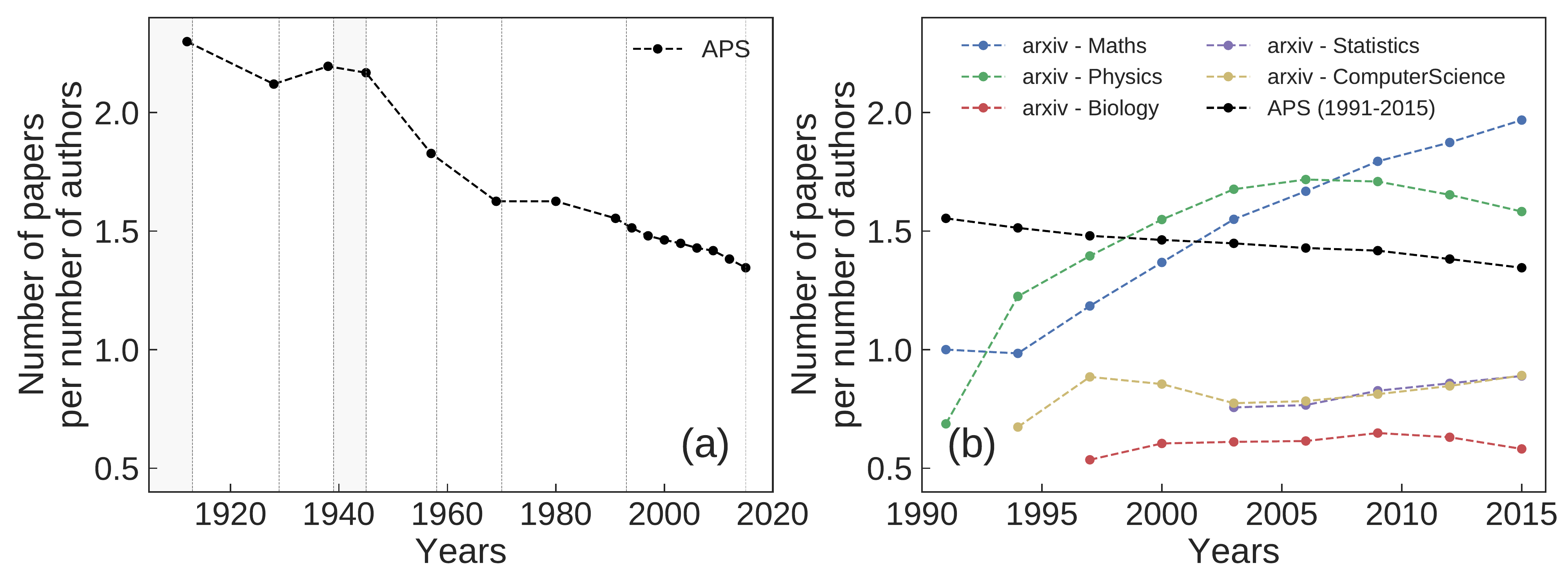}
	\caption{Ratio of the total number of authors to the total number of papers for (a) APS journals and (b) ArXiv repository. The decrease of papers per author is noticeable for APS journals especially after the World War II, until the appearance of large collaborations around the 90s. Maths on ArXiv, on the other hand, has doubled its ratio of total number of papers per total number of authors in the last 25 years.}
	\label{fig:size_rate}
\end{figure*}

The evolution of the size of partitions $U$ and $V$ (i.e. the number of authors and papers in the network, respectively) 
is roughly exponential for all cases, as shown in Figure \ref{fig:network_size}. Such growth behaviour has already been reported in previous studies \cite{huang2008collaboration,martin2013coauthorship,sinatra2015century}. However, in Figure \ref{fig:aps_size} it is possible to see that the total number of authors grows at a faster pace and is approaching the total number of papers. Such behaviour becomes clearer by looking at Figure \ref{fig:size_rate}. In physics, the ratio of the number of papers per author has a noticeable decrease, specifically in the period between the Second World War and the appearance of large collaborations around the 1990s. After the latter, the decrease of such ratio becomes much smaller. Mathematics, on the other hand, has doubled its ratio of papers per author, in the last 25 years. 

This type of comparative analysis between both datasets and across disciplines of the ArXiv dataset will be recurrent throughout this paper. However, a focus will be given to the structural dynamics of the APS network because of its larger time frame.

The remainder of this paper is organised following the topological properties of our interest: in Section \ref{sec:degree} we will look at the evolution of the average degree of the networks as well as the evolution of their degree distributions. We will show how the emergence of large collaborations in physics results in a degree distribution for the top nodes with a much heavier tail than the other disciplines.  Next, we will analyse, in Section \ref{sec:density}, the way density evolves over time in the APS and ArXiv networks. The presence of high degree top nodes --- papers with a large number of authors --- inflates the density of the projected networks in physics (for both datasets). 
Using a generating function approach, we prove that the density in projected networks can present a transition, even if the density level of the original bipartite network remains low. In Section \ref{sec:cycles} we discuss the presence of cycles and, more importantly, small cycles in bipartite networks. Moreover, we discuss the redundancy of top nodes and their effect on the distribution of link weights in the co-authorship network. 
Finally, Section \ref{sec:conclusion4} presents the conclusions of the this study.

\section{Evolution of degree distributions and assortativity}
\label{sec:degree}

As bipartite networks have two sets of nodes, let us recall that it is important to differentiate the degree of nodes in each partition. Top nodes $v \in V$ have degree denoted by $d_{v}$, while bottom nodes $u \in U$ have their degree denoted by $k_{u}$. Degree $d$ represents the number of authors on a paper and the average degree $\langle d \rangle$ in $V$ is the mean number of authors per paper in the network. On the other hand, degree $k$ is the number of papers an author has, and $\langle k \rangle$ is the mean number of papers per author. 

\begin{figure*}[]
	\centering
	\subfigure{\label{fig:apsbipavdegree} \includegraphics[scale=0.42]{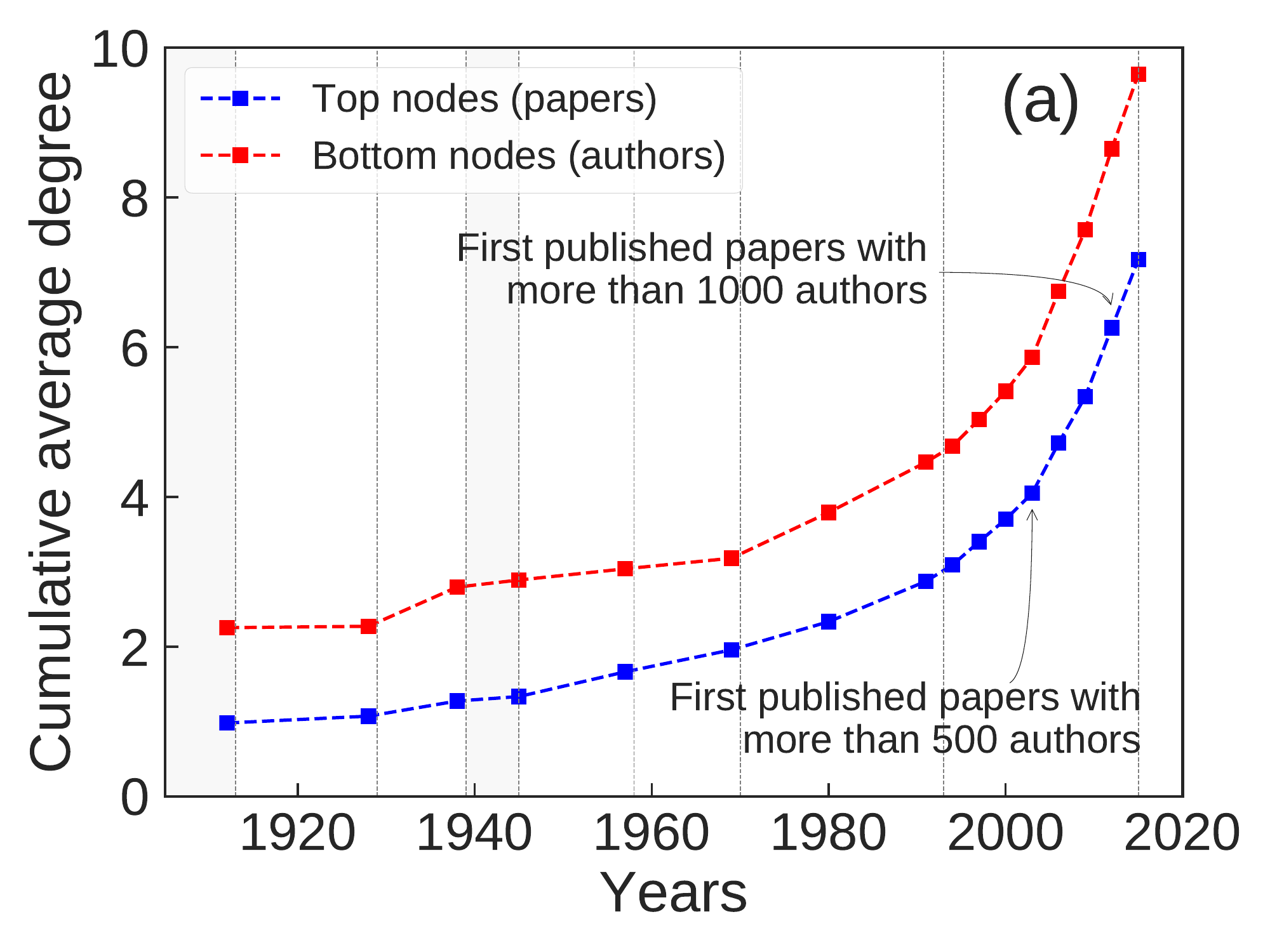}}
	\subfigure{\label{fig:arxivbipavdegree} \includegraphics[scale=0.42]{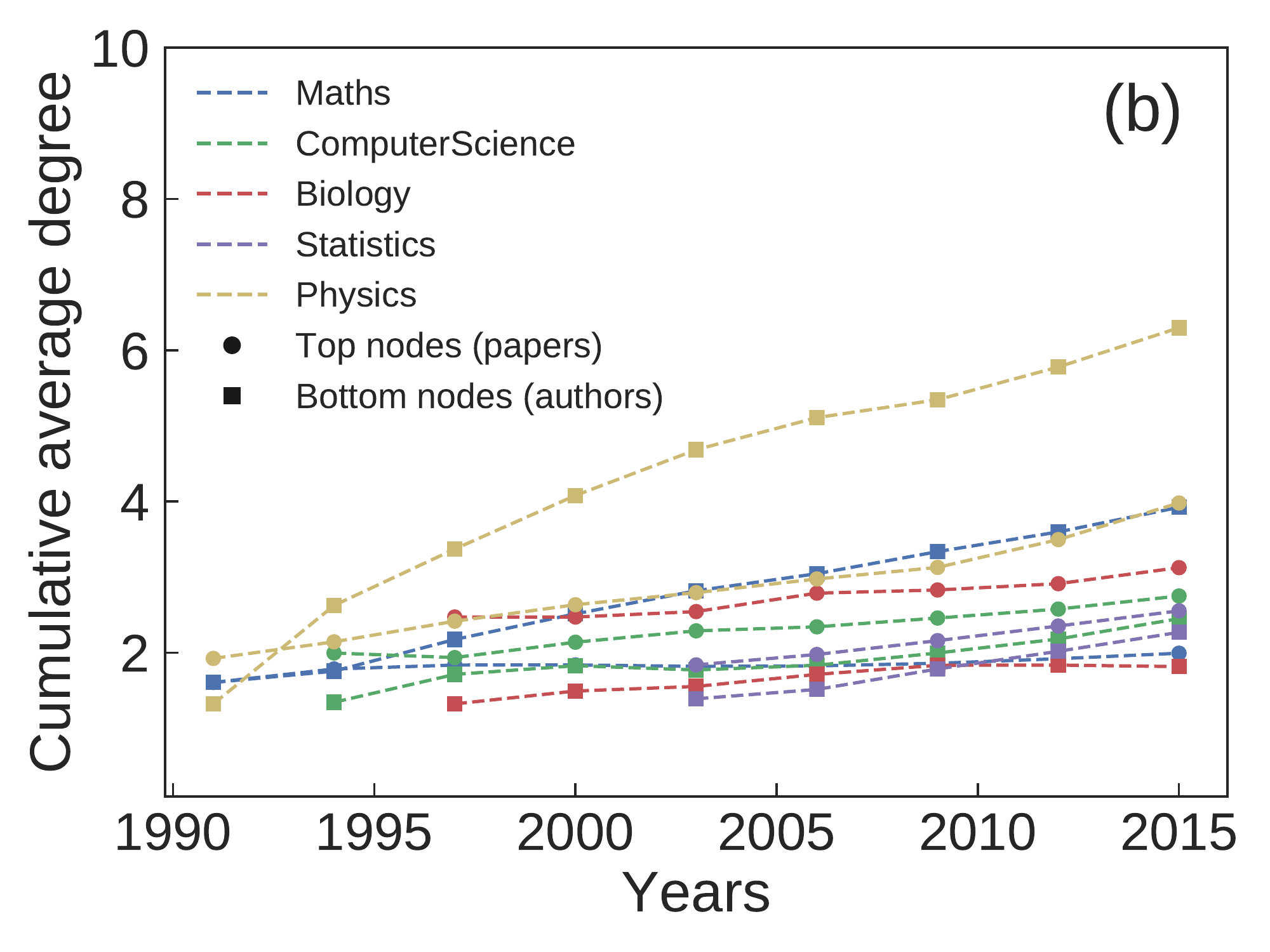}}%
	\caption{Evolution of average degree for bottom and top nodes for (a) APS and (b) ArXiv. While there is a transition for APS journals around the 90s, all ArXiv networks have a steady, roughly linear growth for the bipartite average degrees.}
	\label{fig:bipavdegree}
\end{figure*}

Looking at the evolution of the average degree for top and bottom nodes over time, in the APS network, 
we see that the mean number of papers per author increases roughly linearly \cite{martin2013coauthorship,huang2008collaboration}, with a transition around the 1990s, when the pace steps up substantially (Figure \ref{fig:bipavdegree}). At a first sight, this can be understood as an increase in productivity, as authors are authoring more papers. This is arguable, however, due to the fact that the mean authors per paper is also increasing, at an even faster pace. The relation between the two partitions in a bipartite network, in regard to partition size and node degree, is given by the number of links in the network $|E| = \sum_{u\in U} k_{u} = \sum_{v\in V} d_{v}$.
That is, even though the mean number of papers per author increased, the mean number of authors per paper increased faster. This is also a consequence of what we saw in Figures \ref{fig:network_size} and \ref{fig:size_rate} in the previous section, where the total number of authors is approaching the total number of papers. In other words, a larger number of authors is needed to create the same number of papers. One could state that, in fact, productivity is decreasing in physics. The authors of \cite{sinatra2015century} reached similar conclusions stating that individual productivity is increasing (mean number papers per author), as consequence of collaborative effects (mean number of authors per paper). On the other hand, the overall productivity --- represented by the ratio of the sets size of the bipartite network, ${|V|}/{|U|}$ --- of the field is decreasing as shown in Figure \ref{fig:size_rate} of this paper and in Figure 1d (red line) of \cite{sinatra2015century}. It is important to notice, however, that the results in \cite{sinatra2015century} curiously show this ratio fluctuating above and below $one$ when, in fact, $|V|$ is never smaller than $|U|$ (Figure \ref{fig:network_size} here and Figure 1c in \cite{sinatra2015century}). This indicates a possible mistake that the authors of \cite{sinatra2015century} have made when calculating the ratio ${|V|}/{|U|}$ from their data.

When we look at the ArXiv networks we see a steady, low pace, linear increase of the mean number of authors per paper (Figure \ref{fig:arxivbipavdegree}) in accordance with \cite{wuchty2007increasing}. Thus, it is intriguing that there is, in Figure \ref{fig:bipavdegree}, a striking difference for both mean papers per author and mean authors per paper between peer-reviewed papers (APS journals in Figure \ref{fig:apsbipavdegree}) and non-refereed papers (ArXiv in Figure \ref{fig:arxivbipavdegree}) in physics. The reasons underlying such differences could be explored, however our interest in this work lies in another direction.

\begin{figure}[]
	\centering
	\includegraphics[scale=0.32]{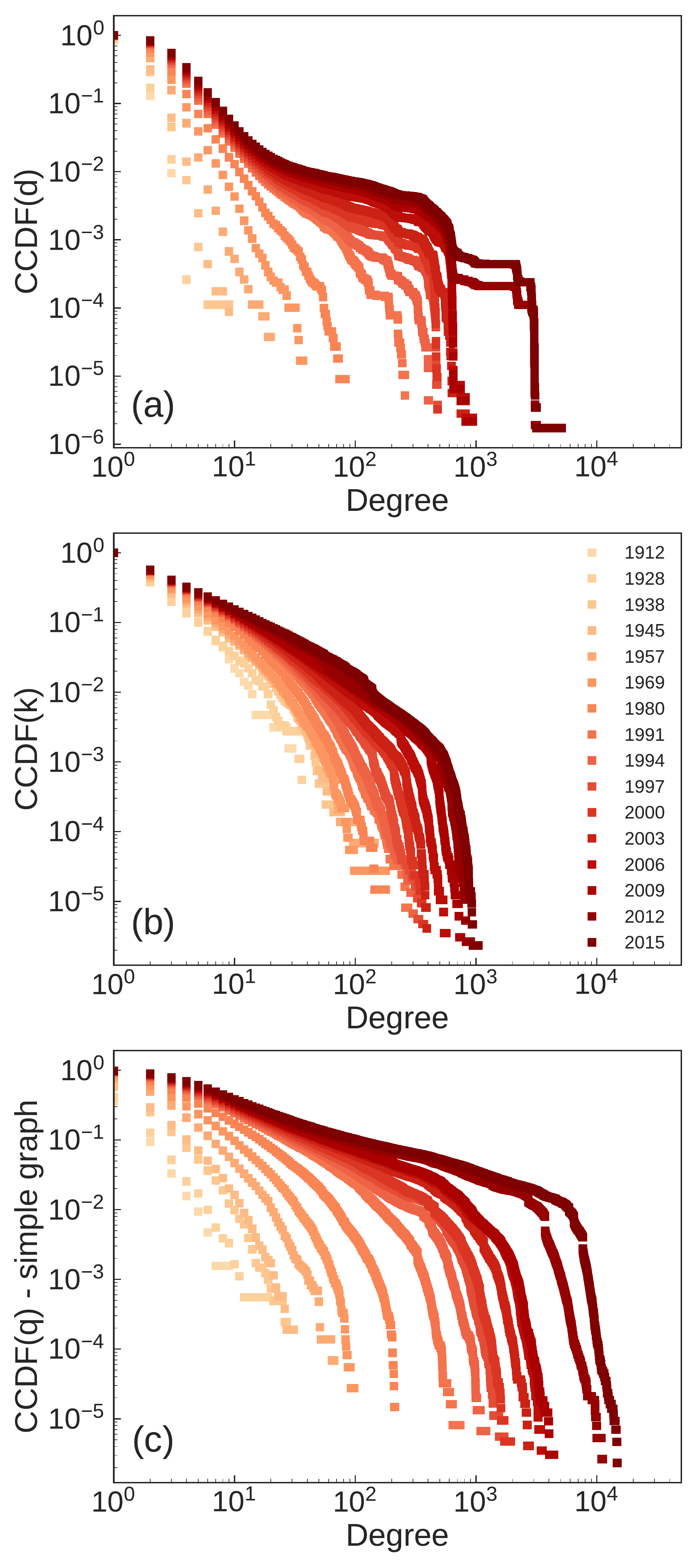}
	
	\caption{Degree distributions for (a) top nodes, (b) bottom nodes and (c) simple (weighted) graph bottom projection. The appearance of large collaborations in physics is notable in the top degree distribution, starting from 1991. Bottom and projected degree distribution has a more well-behaved evolution, roughly keeping the same shape, just shifting to the right as the network grows.}
	\label{fig:degreedistr}
\end{figure}

Looking at the evolution of the degree distributions, 
the increase of the average degree of the papers in the APS network is due to a drastic change in the shape of the top degree distribution. It starts as peaked ending up as a highly skewed distribution (Figure \ref{fig:degreedistr}a).

The early APS network has, from its start until approximately the end of World War II, a mean number of authors per paper of a little over one, increasing from one in 1912 (maximum degree is three) to 1.3 in 1945 (maximum degree is nine). The shape of the top degree distribution is roughly Poissonian. From 1945 to 1970 there is a transition in the top degree distribution, which starts taking the form of a power-law distribution. By then, the sizes of the sets of the network are still of a few tens of thousands of nodes. In 1970 the mean number of authors per paper reaches two, and papers with more than 30 authors already exist. 

A new transition happens again around 1990 when the top degree distribution becomes highly heavy-tailed. At this point, the top average degree --- the mean number of authors per paper --- reaches 2.9 and the maximum degree of a paper is 258; and the network is much larger, with a little over 100,000 authors and almost 200,000 papers published. Since then, the top degree distribution gets more and more extremely right-skewed. By the end of 2015, the mean number of authors per paper reached 7.2, and there was a paper with more than 5000 authors.

\begin{figure}[]
	\centering
	\hspace*{-0.8cm}
	\includegraphics[scale=0.22]{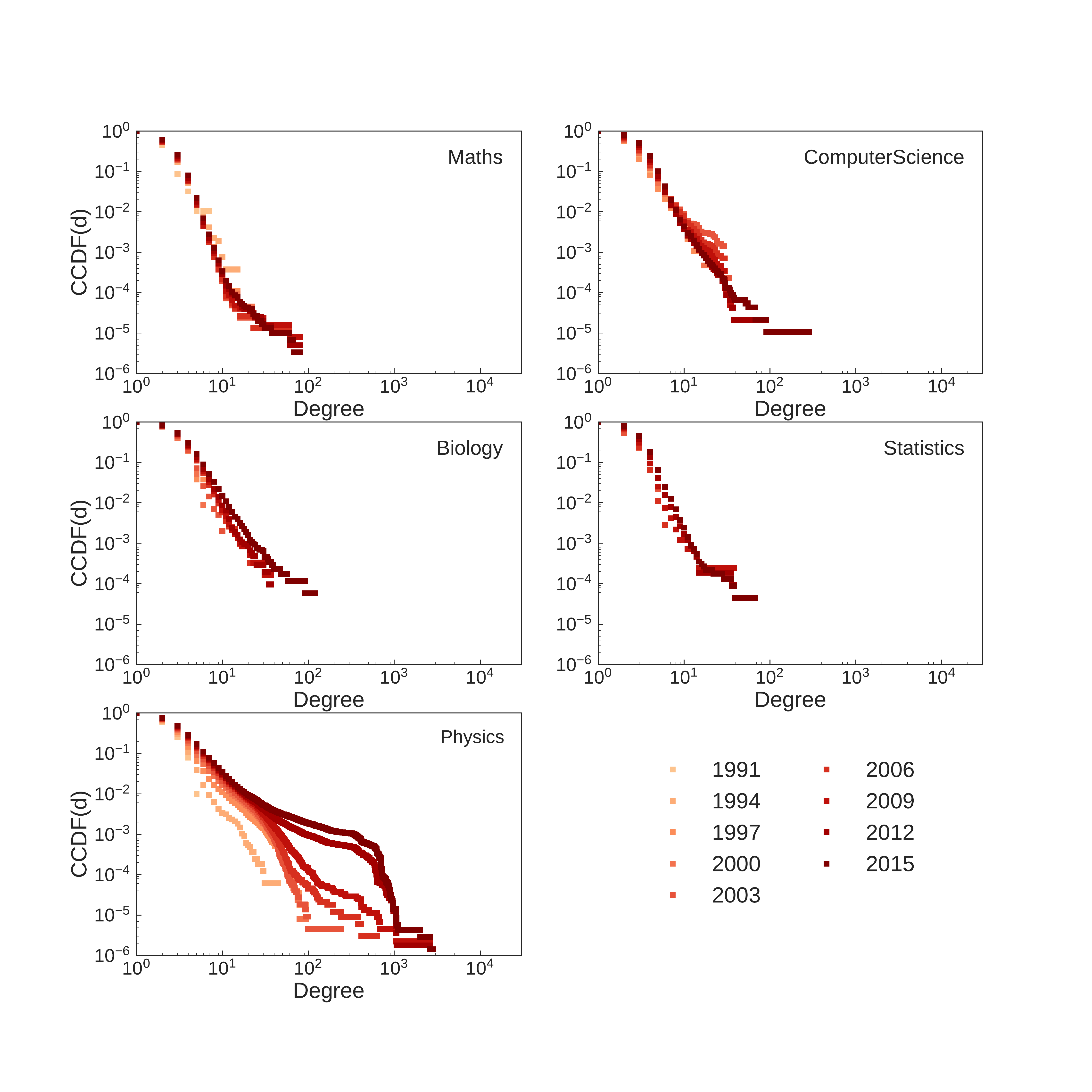}
	\vspace{-0.5cm}
	\caption{Top degree distributions for all five ArXiv disciplines. Physics presents a distribution much more skewed than the others, again due to the large collaborations, mostly after the 90s. While there are papers with thousands of authors in physics, papers with largest degree in other disciplines are around one hundred and couple of hundred (computer science) authors.}
	\label{fig:arxiv_top_distr}
\end{figure}

This behaviour is also seen in physics papers in the ArXiv dataset, but not in the non-physics disciplines. Maths, biology, and statistics have papers with the largest degree of around 100 authors, and computer science has only one paper that exceeds that, with a few hundred authors (Figure \ref{fig:arxiv_top_distr}).

\begin{figure}[]
	\centering
	\hspace*{-0.8cm}
	\includegraphics[scale=0.22]{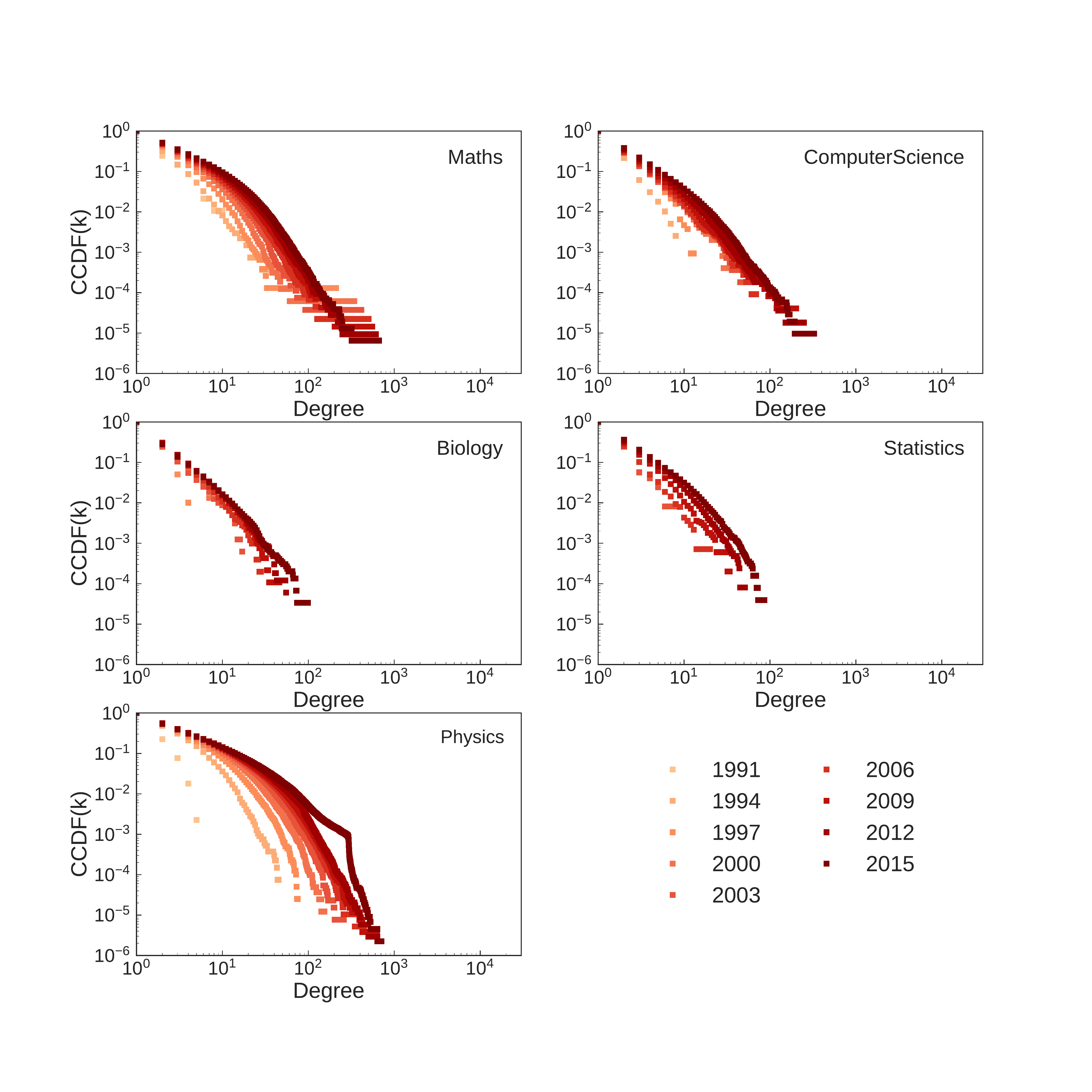}
	\vspace{-0.5cm}
	\caption{Bottom distributions for all five ArXiv disciplines. These are more well-behaved than top degree distributions, mostly because of the time span in which authors can still produce papers. Due to the enormous amount of papers produced by the large collaborations, in physics again we see a certain activity in the bottom degree distribution, as authors that are part of such collaborations become more prolific.}
	\label{fig:arxiv_bottom_distr}
\end{figure}

In contrast, bottom degree distributions (papers per author) do not change so drastically. Figures \ref{fig:degreedistr}b and \ref{fig:arxiv_bottom_distr} show a more well-behaved evolution than that of top degree distributions. The fact that it is difficult for authors to produce a very large number of papers during their career limits changes seen in the bottom degree distribution. In later years, after 2010, some authors appear with around a thousand papers entirely due to the increasing number of publications of large collaborations, for example the ATLAS (part of the Large Hadron Collider particle accelerator, operating since 1992) and the CDF (Collider Detector at Fermilab, operating since 1985) collaboration groups \cite{atlascollaboration,cdfcollaboration}. In the $21^{\mathrm{st}}$ century, many papers are being published with an enormous number of authors, as we described above, making participants of such groups reach high numbers of papers published.

\begin{figure}[]
	\centering
	\hspace*{-0.8cm}
	\includegraphics[scale=0.22]{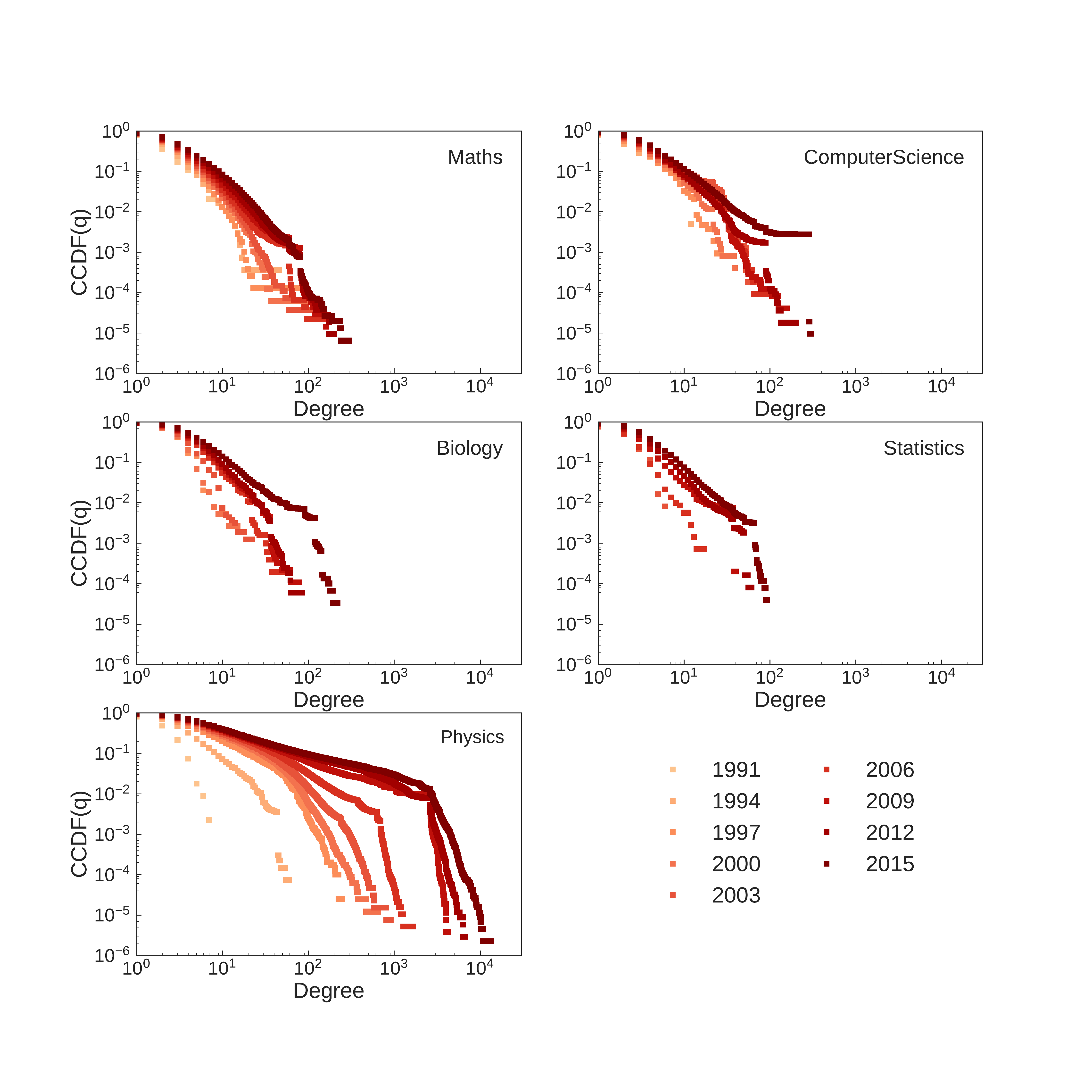}
	\vspace{-0.5cm}
	\caption{Co-authorship degree distributions for all five ArXiv disciplines. All but physics have their projected degree distributions ruled by their bottom distribution of the bipartite networks, which have heavier tail distribution than top nodes \cite{vasques2018degree}. Otherwise, physics top degree distribution, whose tail increases drastically,  creates large complete subgraphs in the co-authorship network. That is, the projected degree distribution is overall ruled by the degree distribution of top nodes.}
	\label{fig:arxiv_projected_distr}
\end{figure}

As we would expect, the degree distributions of the bottom projected physics networks --- of co-authors per author --- is extremely heavy-tailed (Figures \ref{fig:degreedistr}c and \ref{fig:arxiv_projected_distr}) in the recent years. Such behaviour is not a surprise since we have top degree distributions with a heavier tail than the bottom degree distributions \cite{vasques2018degree}, due to the aforementioned large collaborations.

For the projected networks, it is worth to keep in mind that a node $u \in U$ has the same degree $q_{u}$ for both the simple graph ($G_{s}$) and the weighted graph ($G_{w}$) projections, where $q_{u}$ is given by \cite{vasques2018degree}

\begin{equation}
\label{eq:qu}
q_{u} \leq \sum_{j=1}^{k_{u}}(d_{v_j} - 1) \,.
\end{equation}

The equality only holds for the case where no pair of authors have co-authored more than one paper, which is unrealistic. It is worth noticing that the node degree $q_{u}$ in the simple and weighted graph projections is representative of the number of unique co-authors of author $u$ in the co-authorship network. As we will see in Section \ref{sec:cycles}, the value of $q_{u}$ in the inequality of Equation (\ref{eq:qu}) is directly related to the number of cycles of size four that node $u$ is part of in a bipartite network $B$.

For multigraph projections ($G_{m}$), however, the equality of Equation (\ref{eq:qu}) always holds and for this particular case, we refer to the degree of node $u$ as its multigraph degree or node strength, $s_{u}$, such that $q^{\textrm{m}}_{u} = s_{u}$. The strength $s_{u}$  of a node in a multigraph projection is the total number of interactions of an author with its co-authors, i.e. the number of shared publications summed over all the unique $q_{u}$ co-authors of $u$.

On the other hand, the number of shared publications between two nodes $u$ and $u'$ can also be represented as the weight $w_{(u,u')}$ of the link connecting them in a simple weighted graph projection (for other weighted projection methods, see \cite{coscia2019impact}). Then, summing the weights over all $q_{u}$ links ($q_{u}$ co-authors) of author $u$ also gives us the strength of $u$ \cite{barrat2004architecture}, such that $s_{u} = \sum_{j=1}^{k_{u}}(d_{v_j} - 1) = \sum_{u'=1}^{q_{u}} w_{(u,u')}$. 
That is, the strength of $u$ can be given either by its degree in a multigraph projection or by the sum of the weight of its links in a weighted graph projection.

\begin{figure*}[]
	\centering
	\subfigure{\label{fig:apsprojavdegree} \includegraphics[scale=0.42]{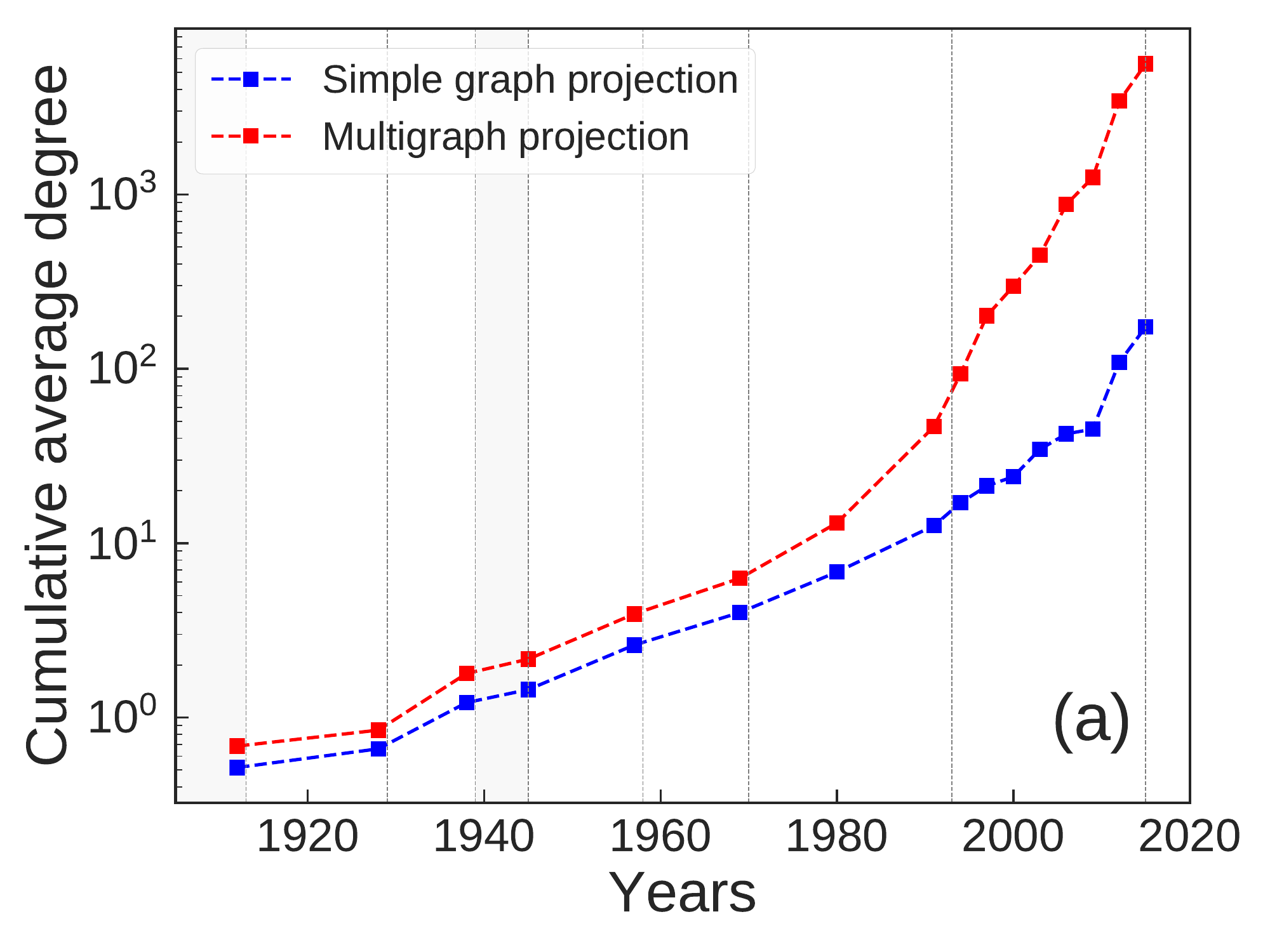}}%
	\subfigure{\label{fig:arxivprojavdegree} \includegraphics[scale=0.42]{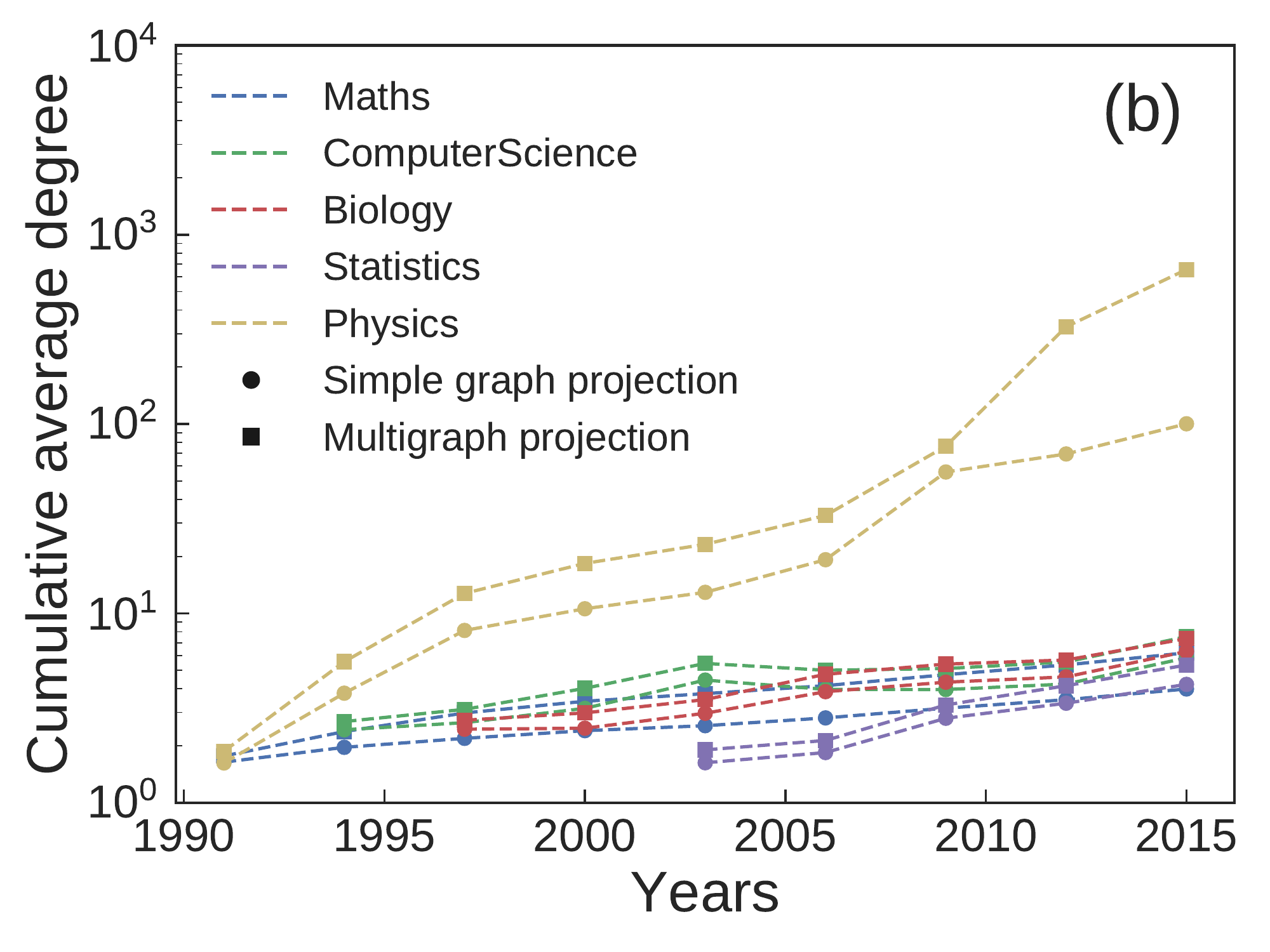}}%
	\caption{Evolution of simple graph average degree (mean co-authors per author) and multigraph average degree (mean multilinks per author) for (a) the APS journals and (b) ArXiv for the disciplines of Physics, Maths, Computer Science, Biology and Statistics. Multigraph average degree is also called average strength. The noticeable difference between physics and the other disciplines is the result of the large collaboration groups.}
	\label{fig:projavdegree}
\end{figure*}

The simple graph average degree (the mean number of co-authors per author) and the average strength (mean shared publications per author) are shown in Figure \ref{fig:projavdegree}. We are able to see, once more, how collaboration in physics has changed significantly for both APS and ArXiv networks over time. This behaviour is quite different from all other four disciplines, in which collaboration patterns have changed very little. The change in physics is due to the rapid increase in the mean number of authors per paper (Figure \ref{fig:bipavdegree}). Every paper $v$ in $B$ induces a clique (complete subgraph) of size $d_{v}$ in $G_{m}$ with 
\begin{equation}
\label{eq:Lv}
|L|_{v} = \frac{d_{v}(d_{v}-1)}{2}
\end{equation}
links. Hence, we can say that for the whole network, the mean strength $\langle s \rangle$ per node in a multigraph projection grows approximately quadratically with the mean number of authors per paper $\langle d \rangle$. Mean co-authors per author $\langle q \rangle$ in $G_{s}$ (and $G_{w}$) does not grow at the same pace because of the inequality of Equation (\ref{eq:qu}). Moreover, we see an increasing difference in the mean number of shared publications and the mean number of co-authors per author in physics, especially after the late 1990s, due to the aforementioned change in the collaboration pattern among physicists. We will discuss it further in Section \ref{sec:cycles}.

We also calculated the evolution of the degree assortativity of the co-authorship networks. At the beginning of the last century, the APS co-authorship network was degree-dissortative. As the number of authors per papers started to increase, the network became neutral by the end of the World War II, and degree-assortative by the 1970s, as shown in Table \ref{table:degree_assort_co}. The co-authorship networks for all ArXiv disciplines also are degree-assortative. This is expected due to the characteristics of the degree distributions and the presence of small cycles (as we will see shortly) in the bipartite networks \cite{vasques2020transitivity}.

\begin{table}[]
	\caption{Evolution of the degree assortativity of the APS and ArXiV co-authorship networks. It is not surprising that all projected co-authorship network are degree-assortative as all bipartite scientific collaboration networks present broad top degree distributions. As seen in \cite{vasques2018degree}, the right-skewed top distribution are more likely to generate degree-assortative projections. The exception is the physics APS network in its early days, when the top distribution was still fairly peaked \cite{vasques2020transitivity}.}
	\label{table:degree_assort_co}
	\centering
		\begin{tabular}{c|c|ccccc}
			\hline
			& APS     & \multicolumn{5}{c}{ArXiv}                                \\
			\hline
			Year & Physics & Physics & CompSci & Maths & Biology & Statistics \\
			\hline
			1913 & -0.218  &         &                 &       &         &            \\
			1929 & -0.111  &         &                 &       &         &            \\
			1939 & -0.021  &         &                 &       &         &            \\
			1946 & 0.019   &         &                 &       &         &            \\
			1958 & 0.134   &         &                 &       &         &            \\
			1970 & 0.376   &         &                 &       &         &            \\
			1981 & 0.48    &         &                 &       &         &            \\
			1991 & 0.619   & 0.516   &                 & 0.723 &         &            \\
			1994 & 0.702   & 0.804   & 0.871           & 0.237 & 0.136   &            \\
			1997 & 0.678   & 0.654   & 0.641           & 0.083 & 0.737   & 1          \\
			2000 & 0.644   & 0.36    & 0.619           & 0.104 & 0.264   & 0.452      \\
			2003 & 0.573   & 0.468   & 0.718           & 0.072 & 0.322   & 0.537      \\
			2006 & 0.522   & 0.775   & 0.592           & 0.51  & 0.519   & 0.147      \\
			2009 & 0.488   & 0.917   & 0.367           & 0.641 & 0.605   & 0.888      \\
			2012 & 0.611   & 0.751   & 0.689           & 0.468 & 0.52    & 0.654      \\
			2015 & 0.59    & 0.524   & 0.978           & 0.312 & 0.879   & 0.64      \\
			\hline
		\end{tabular}
	\vspace{0.5cm}
\end{table}

\section{Densification due high-degree top nodes}
\label{sec:density}

The number of links in the bipartite network, and therefore the density, is directly related to the degree distribution of both sets $U$ and $V$. The density of $B$, $\rho_{\textrm{B}}$, is given by \begin{equation}
\label{eq:rhoB}
\rho_{B} = \frac{|E|}{|U| \cdot |V|}\,,
\end{equation}
where $|U| \cdot |V|$ is the total number of possible links in $B$.
In contrast, the density of the projected network depends on the degree distribution of the top nodes only, and shows no direct relationship with either $|E|$ or $\rho_{\textrm{B}}$, as one could expect. The number of links $|L|$ in a projected network is calculated by summing Equation (\ref{eq:Lv}) over all nodes $v \in V$, such that
\begin{equation}
\label{eq:|L|}
|L| \leq \sum_{v} \frac{d_{v}(d_{v}-1)}{2} = \sum_{v} L_{v}\,.
\end{equation}
The presence of the inequality in the above equation is for the same reasons as of Equation \ref{eq:qu}.

\begin{figure*}[]
	\centering
	\subfigure{\label{fig:apslinks} \includegraphics[scale=0.42]{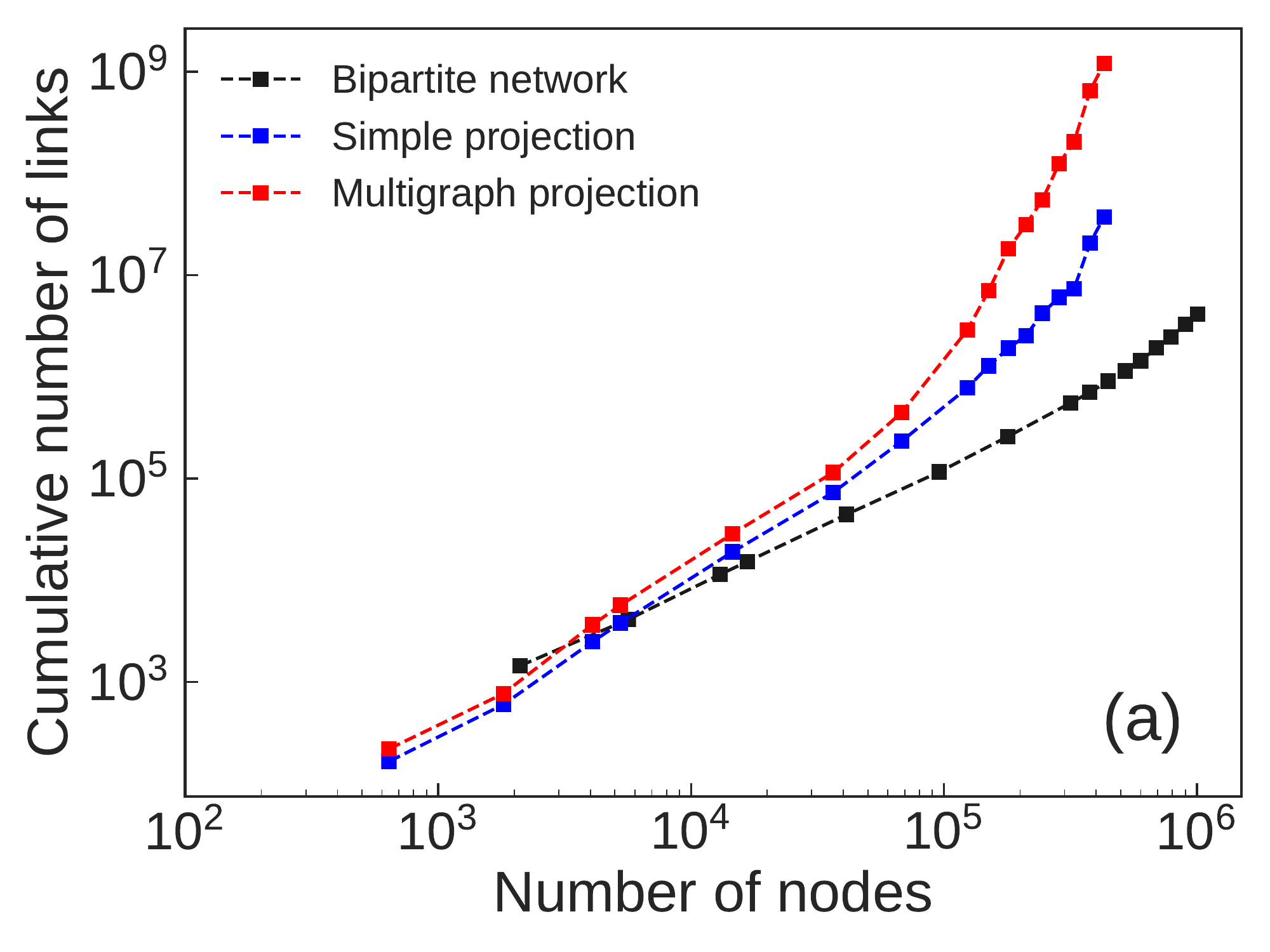}}%
	\subfigure{\label{fig:apsdensity} \includegraphics[scale=0.42]{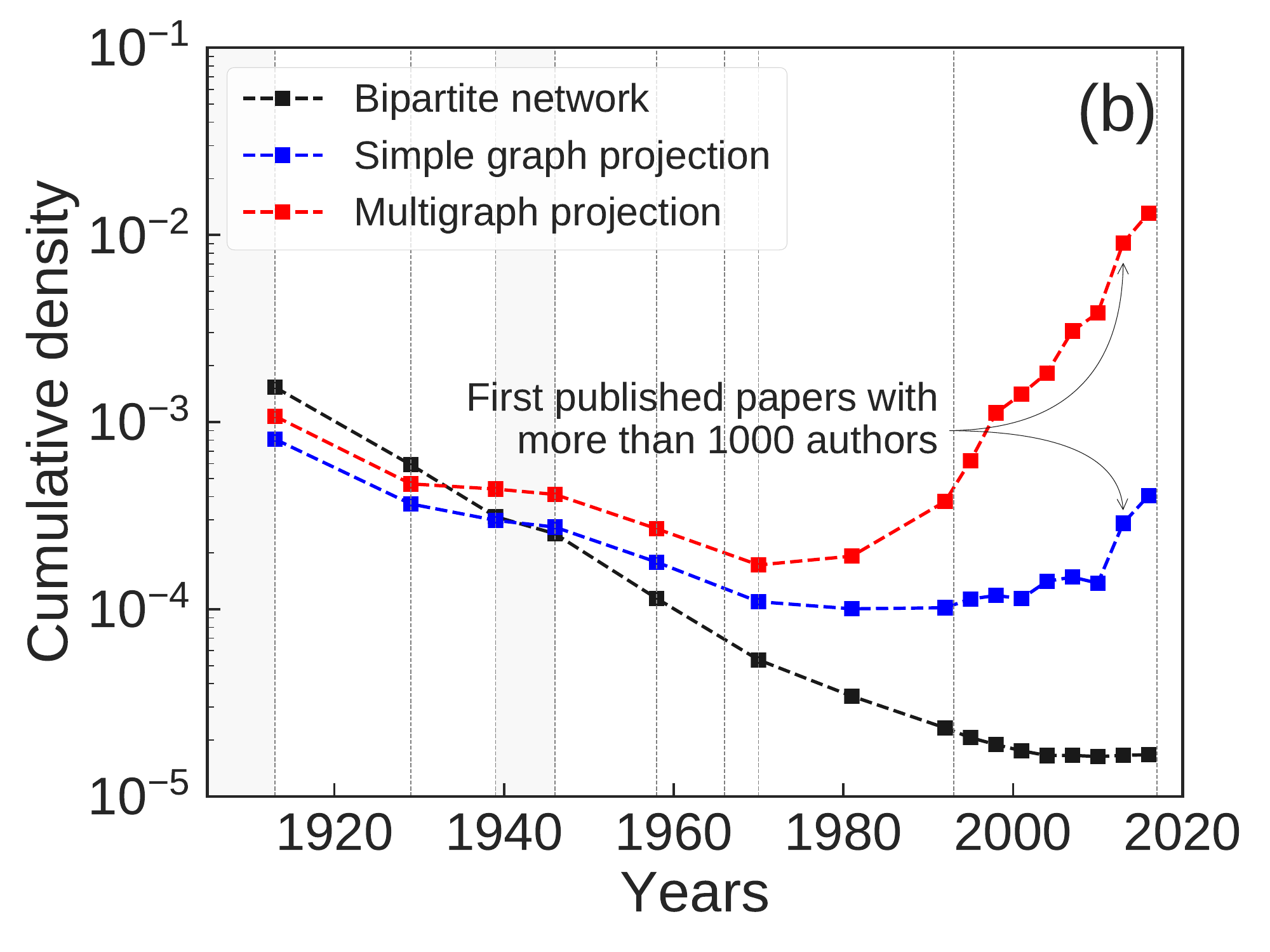}}%
	\caption{Evolution of (a) number of links and (b) density for the APS journals network for the bipartite, simple graph and multigraph networks. Although the density in the bipartite network still decreases, the density of the projections increase due to the heavy tail of the top node distribution.}
	\label{fig:aps_density}
\end{figure*}
\begin{figure}[]
	\centering
	\hspace*{-0.8cm}
	\includegraphics[scale=0.22]{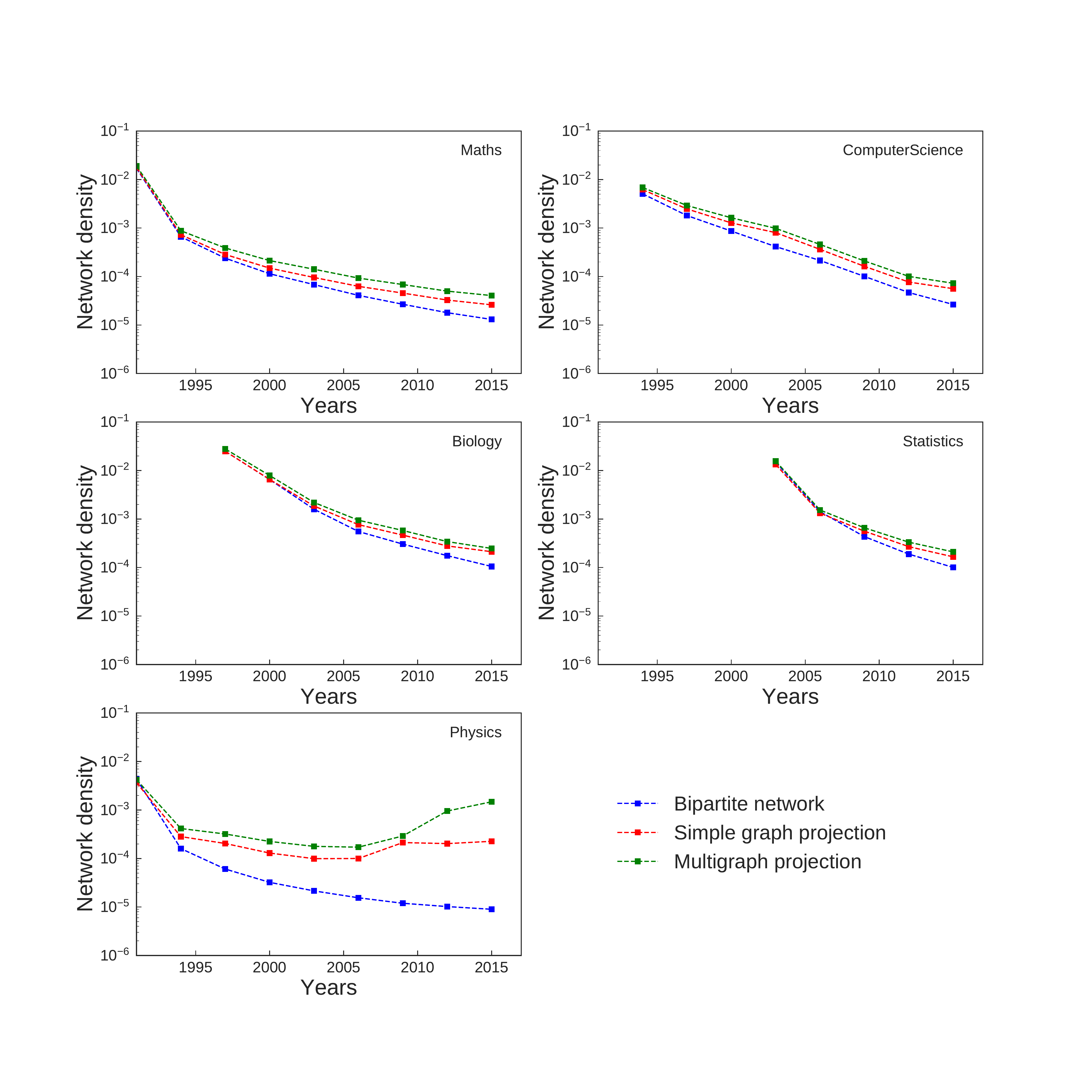}
	\vspace{-0.5cm}
	\caption{Network density for all five ArXiv disciplines. The density decreases for the bipartite network and the projections for every discipline, except physics. One might expect that as soon as the density decreases in the bipartite structure, it will decrease also for the projections. However, as we have seen using Equation (\ref{eq:|L|}), the projection density depends solely on the top node degree distribution. Therefore, even though high-degree physics papers are not enough to increase the bipartite density, they cause the transition observed in the co-authorship network. For the bipartite network, the contribution of the degree of each top node $v$ is linear, while for the projection the contribution is quadratic.}
	\label{fig:arxiv_net_density}
\end{figure}

Interestingly, we note a transition of the density of the APS projected networks during the 1980s (Figure \ref{fig:apsdensity}). This happens as the degree distribution of top nodes changes from a peaked to a heavy-tailed form, as discussed in the previous section. In the ArXiv physics co-authorship network, a similar behaviour is seen. However, as other disciplines do not present large collaborations --- meaning that the top degree distribution is much less skewed --- their ArXiv networks do not show any densification (Figure \ref{fig:arxiv_net_density}). In addition, it is also worth noticing the jump in the density level when very high-degree top nodes (papers with more than 1,000 authors) appear in the network. From Equation (\ref{eq:Lv}), a node $v$ with degree, let us say, $d_{v} = 1000$ will induce a clique (complete subgraph) of size 1,000. Such a clique will have, by itself $L_{v} = 499,500$ links. This indicates that research in physics has drastically changed, especially for experimental physics. Many groundbreaking findings depend now on the existence of large collaboration groups involving researchers with a wide range of skills.

Let us take a closer look at the analytical explanation for this. The density of the projected network is given by
\begin{equation}
\rho_{\textrm{G}} = \frac{2|L|}{|U|(|U|-1)}\,,
\end{equation}
where $|U|$ is the number of bottom nodes. $|L|$ is given by Equation (\ref{eq:|L|}) or, more conveniently, by
\begin{equation}
|L| \leq \sum_{d=1}^{d_{max}} |V| P_{t}(d) \frac{d_{u}(d_{u}-1)}{2}
= \frac{|V|}{2} \sum_{d=1}^{d_{max}} P_{t}(d) d_{u}(d_{u}-1)\,,
\end{equation}
where $P_{t}(d)$ is the top degree distribution and $|V|$ is the number of top nodes. Now we must recall the generating function approach for the analysis of networks \cite{newman2001random,wilf2013generatingfunctionology,vasques2018degree}. The term inside the sum, in the equation above, is the second derivative of the generating function for $P_{t}(d)$, for $x=1$, such that 
\begin{equation}
|L| \leq \frac{|V|}{2} f''(1)\,.
\end{equation}
In order to clarify, we work out a simple example, much simpler than the APS network. Consider that we have now a network with $|U|$ and $|V|$ growing at the same rate, one new node for each time step, such that $|U|=|V|=t$. We assume that the top distribution is Poissonian, meaning that, as the network grows large ($\lim_{t\to\infty}$), the generating function for the distribution is given by
\begin{equation}
f(x) = e^{\langle d \rangle (x-1)}\,,
\end{equation}
and the second derivative of $f(x)$ is
\begin{equation}
f''(x) = \langle d \rangle^{2} e^{\langle d \rangle (x-1)}\,.
\end{equation}
Therefore, for a degree distribution of top nodes in $B$ following a Poisson distribution, with the set of nodes $|V|$ growing as mentioned above, the density of the projected network is expected to be
\begin{equation}
\label{eq:rhofic}
\rho_{\textrm{G}} \leq \frac{\langle d \rangle^{2}}{(t-1)}\,.
\end{equation}

It is worth noticing that, for the case of $\langle d \rangle$ being constant as the network grows, $\rho_{\textrm{G}}$ will never have a turning point like that observed in the APS co-authorship network. For the transition to happen in our fictitious network, $\langle d \rangle$ must be a function of time such that, in Equation (\ref{eq:rhofic}), it grows faster than the denominator. 
That is obviously not the case of the physics co-authorship network, where $\langle d \rangle$ grows in a much slower pace than the size of the set of nodes $V$. 

Two conditions are in hand. First, the number of links in the projected network grows proportionally to the second moment of the generating function of the top degree distribution. Second, the average degree grows at a much slower pace than the number of nodes in the APS network. In this scenario, the only way that the projected density will start to increase is if the top node degrees follow a heavy-tailed distribution. That is, high degree top nodes inducing big cliques are the reason for the transition in the density level of the projected (co-authorship) network, even though the density of the bipartite network remains low. 

\section{Increasing frequency of small cycles}
\label{sec:cycles}

In the previous two sections, we have seen that the degree distributions of a bipartite network play an important role in the structure formation of projected networks. Both the degree distribution and the density of projections are affected by the shape of the bipartite degree distributions. However, something else is also affecting the structure of projected networks as we see that the difference between the mean number of co-authors per author $\langle q \rangle$ and the mean number of multilinks per author $\langle s \rangle$, in Figure~\ref{fig:apsprojavdegree}, and the difference between the densities of the simple graph and the multigraph projections, in Figure \ref{fig:apsdensity}, are increasing.

As we stated before, differences between the degree $q_{u}$ of $u$ in a simple graph and its strength $s_{u}$ in a multigraph projection is due to the inequality in Equation (\ref{eq:qu}), which is a consequence of four-cycles in the bipartite network \cite{vasques2019bipartite}. The increase in the number of collaborations we see in physics, especially with the rise of large collaboration groups, drastically increases the number of four-cycles in the network. 

Although larger cycle sizes are important in the formation of new smaller cycles in $B$, as we will see later in this section, it is important to make a clear distinction for cycles of size four. They are independent of cycles of other sizes (i.e. they are not created as a consequence of bigger cycles), and are the representation of recurrent pairwise collaborations. Hence, there is significant growth of the frequency of such cycles as new large collaboration groups emerge in physics. The number of four cycles generated by a group of $N$ collaborators with $n$ published papers is given by
\begin{equation}
\textrm{\# four-cycles} = \frac{n(n-1)}{2} \binom{N}{2} \,,
\end{equation}
while the number of fundamental cycles \cite{paton1969algorithm} --- the minimum set of cycles with which all other cycles can be created in the network --- is $(N-1)(n-1)$.

\begin{figure}[]
	\centering
	\includegraphics[scale=0.40]{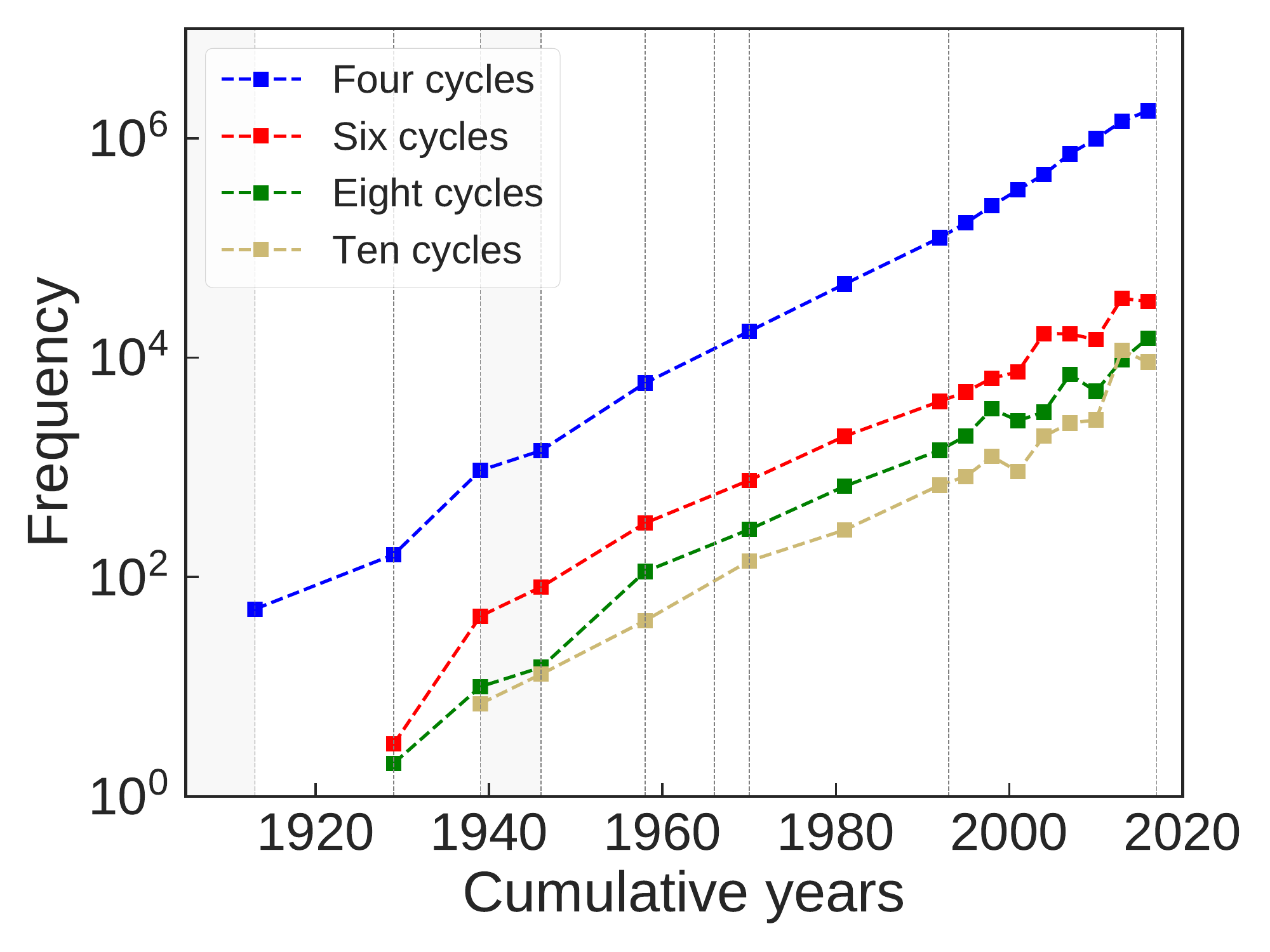}
	\caption{Evolution of the number of small cycles, from size four to size ten, in the APS bipartite network. Four-cycles represent the recurrence of collaboration between pairs of nodes. Six-cycles, in turn, tell us the level of transitivity of the co-authorship network. Transitivity would be hard to calculate through projected networks, without the information of the presence of six-cycles in the bipartite network, as triangles are also induced by top node degrees, for every $d_{v}\geq3$.}
	\label{fig:small_cycles}
\end{figure}

\begin{figure}[]
	\centering
	\hspace*{-0.8cm}
	\includegraphics[scale=0.22]{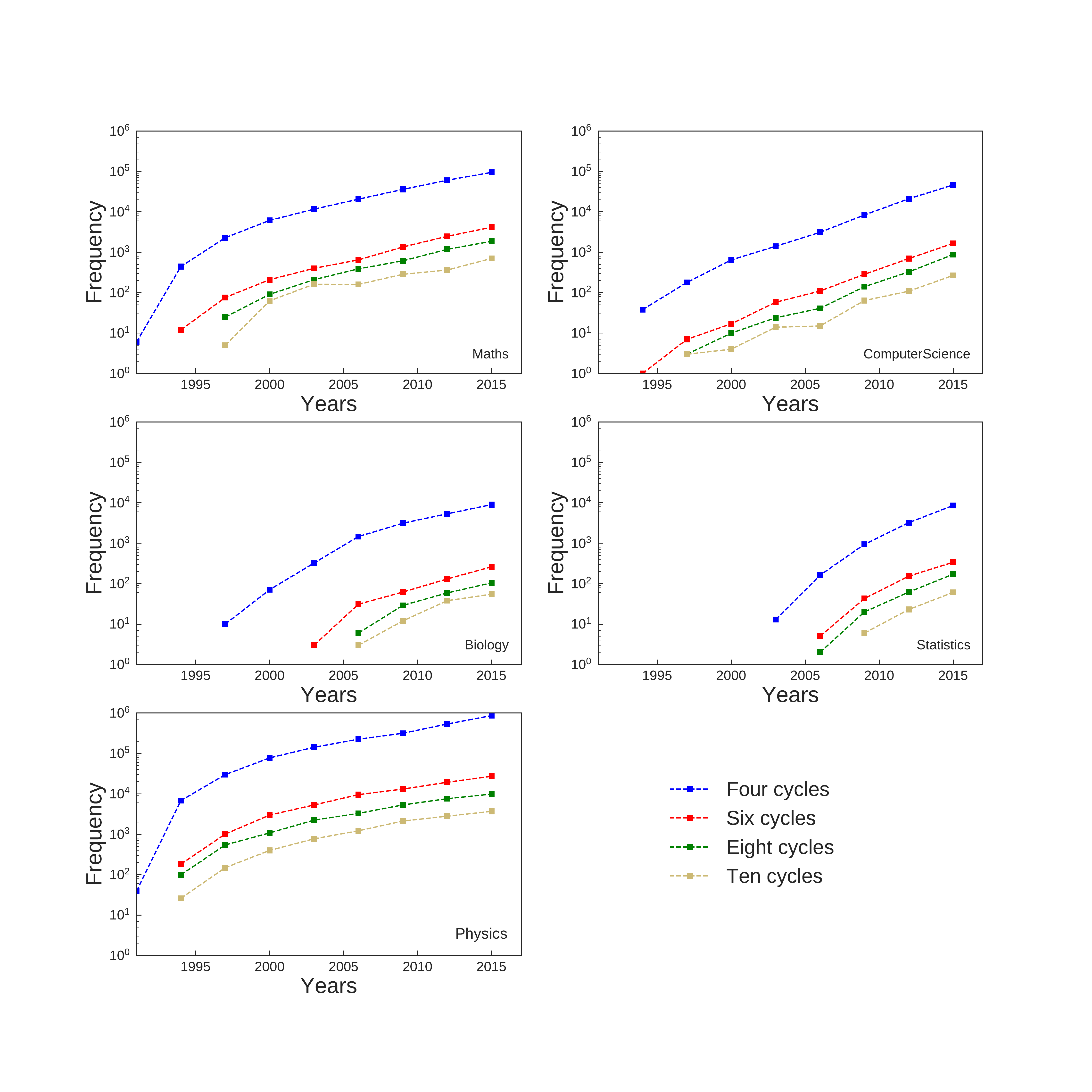}
	\vspace{-0.5cm}
	\caption{Evolution of the number of small cycles, from size four to size ten, in the ArXiv bipartite network. Apart of the great differences in the degree distributions between physics and the other disciplines, the structure and proportionality of the small cycles seem to follow the same pattern.}
	\label{fig:arxiv_small_cycles}
\end{figure}

Four-cycles are present in the network from the beginning, but as the network grows, cycles of bigger sizes start to appear. In Figures \ref{fig:small_cycles} and \ref{fig:arxiv_small_cycles} we see the evolution of the number of small cycles present in the cycle basis of the APS and ArXiv bipartite networks, respectively. 

The evolution of small cycles seems to follow a standard pattern across all disciplines and also for the APS network, unlike what we saw for the degree distributions. Moreover, such a pattern reveals how these small cycles are affecting the structure of bipartite networks and their projections. First, four-cycles show recurrence of interactions between pairs of nodes meaning more redundancy and link weight, as we will see shortly. In other words, authors tend to collaborate repeatedly with past collaborators.  Second, six-cycles are another factor that contribute to transitivity in a network, that has an original bipartite structure. This is due to the fact that projected networks are full of triangles induced by top nodes, for every node $v$ such that $d_{v}\geq3$. The presence of many six-cycles and the high level of transitivity shows that the co-authorship networks have a high level of information flow \cite{schilling2007interfirm}.

\begin{figure}[]
	\centering
	\includegraphics[scale=0.28]{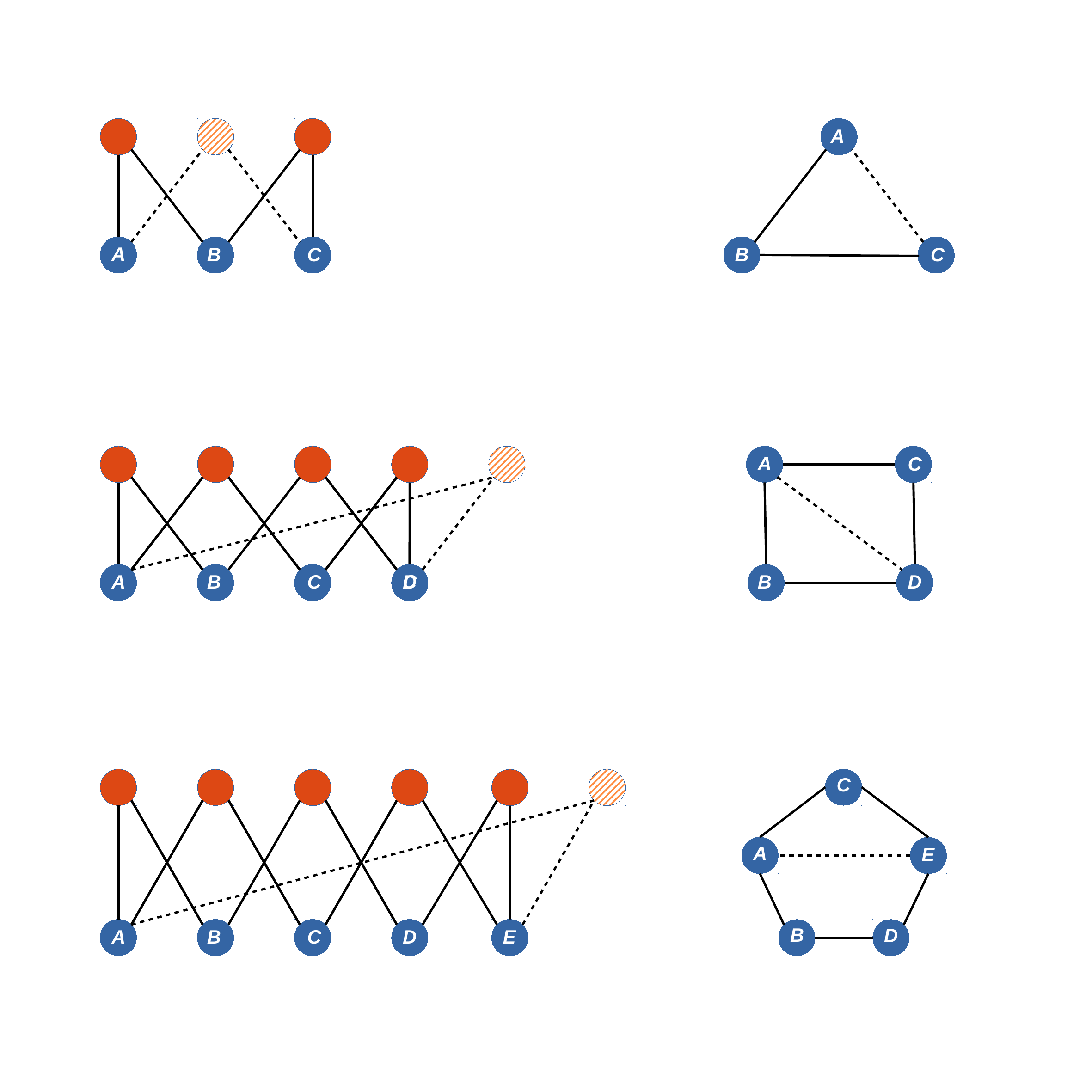}
	\caption{Schematic of six-cycles being created from larger cycles, with the addition of one new publication. The creation of a new six-cycle represents the transitivity increasing in the bipartite network and triadic closure in the projection. 
	}
	\label{fig:cycles_formation}
\end{figure}

Although cycles of size six can also be formed independently, they are mainly formed as a consequence of larger cycles. 
In Figure \ref{fig:cycles_formation} we show a schematic where six-cycles are formed either independently of, or when they are embedded in, larger cycles like eight-cycles and ten-cycles.

Due to the presence of four-cycles in the bipartite networks, simple graph and multigraph projected networks have distinct values of density. This is related to the notion of redundancy, $rc_{v}$, given by \cite{latapy2008basic,vasques2019bipartite}
\begin{equation}
	 rc_{v}= \frac{|\{\{u,u'\} \subseteq N(v), \exists v' \neq v, (v',u) \in E \textrm{ and } (v', u') \in E\}|}{L_{v}} \,.
\end{equation}
As a reminder, $rc_{v}$ is the fraction of the total number of links $L_{v}$ that $v$ induces in $G_{s}$ that remain in the network even if $v$ is removed from $B$. Such links remain present because they are induced by other top nodes.

\begin{figure*}[]
	\centering
	\subfigure{\label{fig:apsredundancy} \includegraphics[scale=0.33]{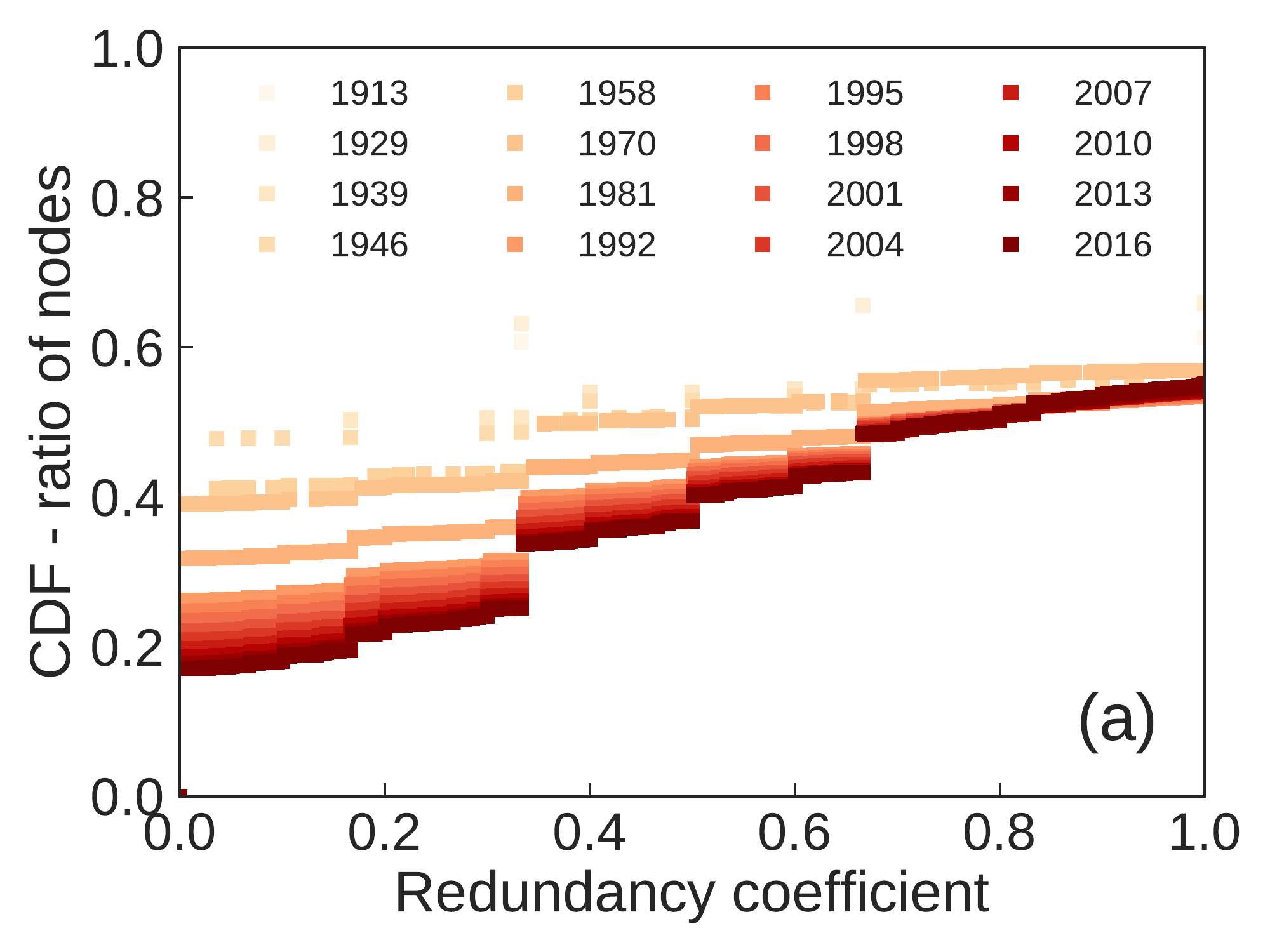}}%
	\subfigure{\label{fig:apslinkweight} \includegraphics[scale=0.33]{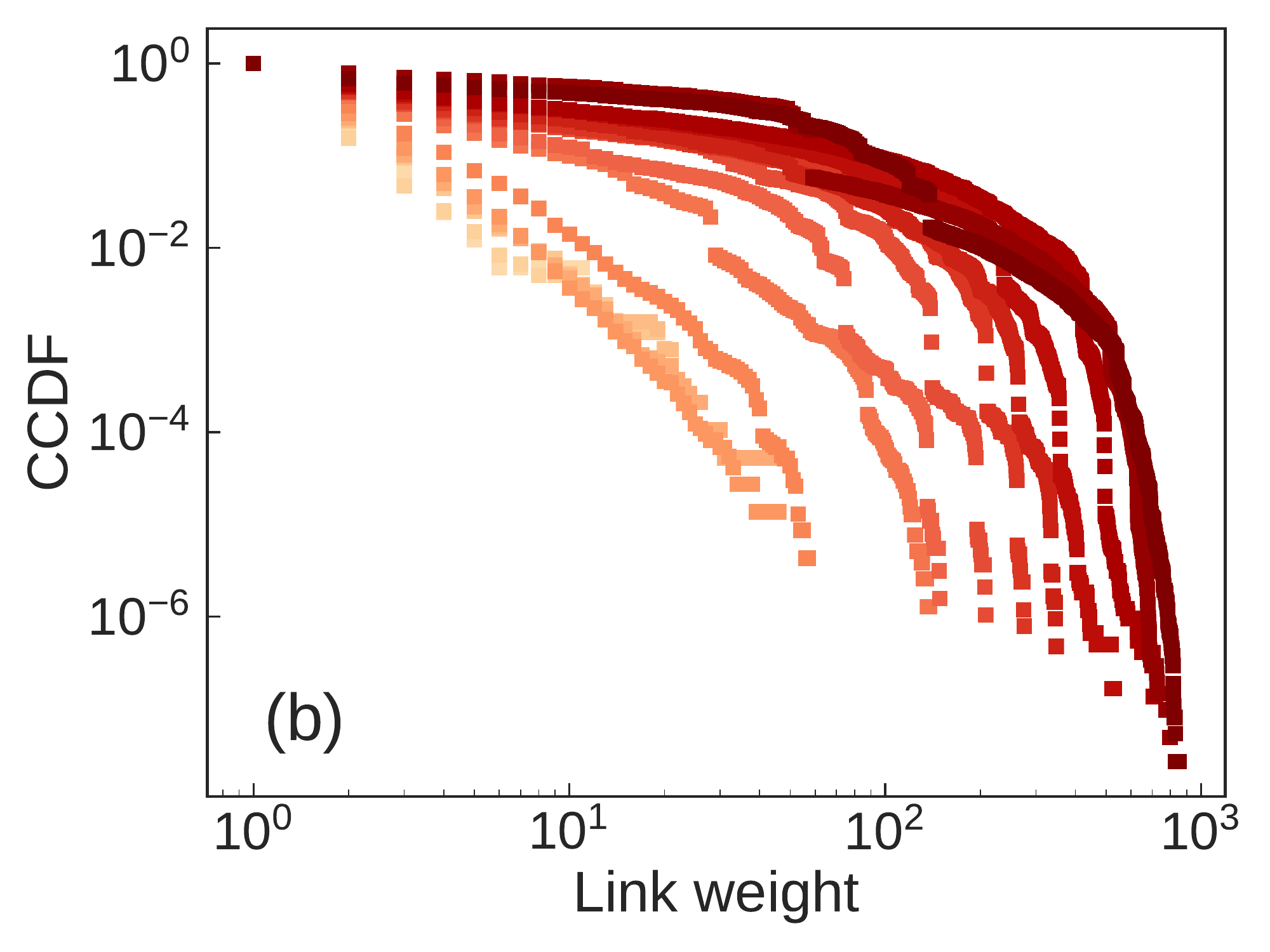}}%
	\caption{Evolution of (a) redundancy and (b) link weight distribution for the APS journals network for the top nodes of the bipartite network and for the weighted graph network, respectively. As redundancy grows bigger, the link weight distribution is pushed to the right.}
	\label{fig:aps_red_weight}
\end{figure*}

In other words, the more redundant top nodes present in the bipartite network, the bigger the difference between the simple graph degree distribution and the strength distribution of projected networks. Such differences are reflected also when we look at the distribution of link weights in a weighted projected network. As stated before, the strength of the node is the sum of the weights of all its links. In Figure \ref{fig:aps_red_weight}, we see the increase of redundancy as the network grows (Figure \ref{fig:apsredundancy}) leading to the increase of the tail of the link weight distribution (Figure \ref{fig:apslinkweight}). The heavy-tailed nature of the link weight distributions shows, once more, the preference that authors have for collaborating with past collaborators. It is worth noticing though, that many heavy-weight links are a result of large collaborations, with some pairs of nodes collaborating in the enormous number of around 800 publications. 

\section{Conclusion}
\label{sec:conclusion4}
In this study, we use two datasets, namely APS and ArXiv, to create six different scientific bipartite networks --- physics from APS and physics, mathematics, computer science, biology and statistics from ArXiv --- and their respective projections, the so-called co-authorship networks. Our goal was to investigate the causalities between the evolution of topological properties of the bipartite networks and the structure of the projections.

The first property that we looked at was degree distributions. With time, the top degree distributions of the physics networks (both APS and ArXiv) became highly skewed, due to the emergence of large collaborations. The other disciplines, however, do not present such collaborations and therefore their top degree distributions do not show tails as heavy as those in physics. The bottom degree distributions do not show any significant change in their shape over the years. They just get shifted to the right as new authors appear in the network and existing authors increase their number of publications. 

As a first result of the behaviour of the evolution of top and bottom degree distributions in the APS network, the mean number of authors per paper and mean papers per author increases steadily, roughly linearly throughout the $20^{\mathrm{th}}$ century, until the 1990s, when the pace steps up substantially, and the mean values more than double in the last 25 years. Interestingly though, we do not see the same behaviour for the ArXiv physics network. The mean number of authors per paper present a significant difference between the peer-reviewed journal and the pre-print repository. One hypothesis for this could be early-career researchers including senior authors in their peer-reviewed papers as a way to gain visibility. However, hypotheses of such a nature are not in the scope of this work. Therefore we will not expand on the matter, and will leave the question open for future endeavours. Secondly, the top nodes with high degree induce large cliques in the physics co-authorship networks, again making their degree distributions (distribution of co-authors per author) significantly different from the other disciplines. Physics co-authorship networks have extremely heavy-tailed top degree distributions when compared to the others with a much higher value of co-authors per author.

A third consequence of the large collaborations in physics, and related to the number of co-authors per author in the co-authorship networks, is the evolution of the density of latter. While one might think that the density of the co-authorship network depends on the density of the bipartite network, we found that this is not true. As soon as many new high-degree top nodes appear in the network, the density of the projection increased considerably, even with the density of the bipartite network keeping low values. We analytically showed the causality between top degree distribution and the densification of the projected network, by making using of the generating function formalism.

We have shown that the number of small cycles grows roughly exponentially for all scientific networks. Four-cycles are created in every newly published paper with the presence of a pair of authors that have co-authored previously. By performing a more detailed analysis of the APS network, we have seen that the distribution of such recurrent co-authorships --- the distribution of link weights in the projected network --- is heavy-tailed, shifting more and more to the right over time. Furthermore, when projecting, eight- and 10-cycles induce squares and pentagons, respectively. Any new co-authorship between any pair of nodes present in the structure and not already connected will create a triangle or, in other words, a new six-cycles is created in the original network. Therefore, six-cycles can appear independently or embedded in eight- and 10-cycles, representing transitivity and triadic closure in the co-authorship networks. 

The implications of our results become clearer when we compare co-authorship networks with an another type of collaboration network. In a previous work, we have analysed a co-patenting network --- where institutions form alliances for innovating \cite{vasques2019bipartite}. 

Although the reasons for which the collaborations take place are different between the co-patenting and co-authorship networks, they display similarities in some features. Institutions and authors both tend to collaborate with past collaborators, with all networks presenting large numbers of four-cycles. Likewise, transitivity is frequent in both co-patenting and co-authorship networks revealing high levels of flow of information. On the other hand, the co-patenting network does not show degree assortativity, while the co-authorship networks do, especially due to the broad top degree distributions in their underlying bipartite networks. 


We conjecture that the difference in the degree-assortativity is that small institutions may be able to fill specific technological gaps of large corporations, such that collaboration for patents do not involve many institutions --- to avoid information spreading to competitors --- and are between those with different collaboration capacity (i.e. large corporations are able to collaborate with many small institutions, while the latter do not have such capacity due to their limited resources). Authors of scientific publications are not so concerned with competition, instead choosing to maximise their productivity and the number of papers that they are listed on, as authors. Therefore, prolific authors with high capacity of collaboration tend to collaborate with other prolific authors with a similar capacity. In this case, information is not trapped and can reach longer distances in the network.




\bibliographystyle{unsrt}  
\bibliography{scinets}

\end{document}